\documentclass{llncs} 
\usepackage{graphicx,color}
\usepackage{amssymb}
\usepackage{amsmath}
\usepackage{xspace}
\usepackage{focus}
\usepackage{bbding} % check marks
\usepackage{pifont} % arrows
\usepackage{stmaryrd} % arrows
\usepackage{mdwlist} %list
\usepackage[usenames,dvipsnames]{xcolor}
\usepackage{enumerate}
\usepackage{todo}

\usepackage{url}

\newcommand{\tbf}[1]{\textbf{#1}}

\newcommand{\dto}{\Rightarrow}

\newcommand{\focust}{\textsc{Focus}$^{ST}$}

\begin{document} 
		\title{Spatio-temporal features of \focust \\ 
}
\author {Maria Spichkova}
\institute{RMIT University, Melbourne, Australia\\ \email{maria.spichkova@rmit.edu.au}  
}
\maketitle 

\begin{abstract}
In this technical report we summarise the spatio-temporal features and present the core operators of \focust specification framework. 
We present the general idea of these operators, using a Steam Boiler System example to illustrate how the specification framework can be applied.

\focust was inspired by \Focus\cite{focus},
a framework for formal specification and development of interactive systems. In contrast to \Focus, 
\focust is devoted to specify and to analyse spatial (S) and timing (T) aspects of the systems, 
which is also reflected in the name of the framework: the extension $^{ST}$ highlights the spatio-temporal nature of the specifications. 
\end{abstract}

%============================
\section{Introduction}

This report summarises our current work on specification and development of safety-critical systems focusing on the spatio-temporal aspects. 
We present here the core operators and features of the \focust specification framework, with the goal to provide a basis for further spatio-temporal analysis of the system properties as well as an interface for the further connection of the \focust\ to other frameworks and techniques. \focust has a high potential for further extensions. For example, Alzahrani et al. \cite{alzahrani2016spatio} proposed spatio-temporal meta-model for property based testing, where the proposed instantiations of a generic modelling language can be TLA+ and \focust.

\focust was inspired by \Focus\cite{focus},
a framework for formal specification and development of interactive systems. 
In our earlier work we developed an extension to \Focus with the goal to cover the timing aspects 
and to provide a specification and verification methodology \emph{\Focus on Isabelle}~\cite{spichkova,spichkova2013we}. 
\emph{\Focus on Isabelle} allows  to specify system in a way that makes further proofs of their properties easier and scalable. 
The methodology also provides a schematically translation to a Higher-Order Logic representation for Isabelle/HOL, an interactive semi-automatic theorem prover Isabelle~\cite{npw}. This allows us to verify the  system properties (that are specified in the extended \Focus framework) using Isabelle/HOL semi-atomatically,
also applying its component Sledgehammer~\cite{Sledgehammer} to benefit from automated sub-goals verification using
first-order automatic theorem provers (ATPs)
and satisfiability modulo theories (SMT)
solvers  in interactive proofs. 

Other advantages of \emph{Focus on Isabelle} are (1) a well-developed theory of composition; 
(2) representation of processes within a system~\cite{spichkova_processes}; 
(3) feasibility shown on a number of auto motive case studies, also formalising the core aspects of the FlexRay communication protocol \cite{spichkova2006flexray,kuhnel2006upcoming,kuhnel2006flexray,feilkas2011refined,feilkas2009top}.

\focust 
was developed on the basis of the extended \Focus, while  giving a special attention to 
\begin{itemize}
\item
 the human factor analysis within formal methods~\cite{hffm_spichkova,Spichkova2013HFFM,spichkova2015human} to increase the readability and understandability of the formal specifications;
\item
spatio-temporal aspects of the  safety-critical systems,  which is also reflected in the name of the framework: the extension $^{ST}$ highlights the \emph{S}patio-\emph{T}emporal nature of the specifications.
\end{itemize}
As result, it allows us to create concise formal specifications that are easy to read and to understand.

The \focust\ specification layout is similar to \Focus (which layout was inspired by Z specification language, cf. \cite{Spivey_88,Spivey_92}), 
but it has many new features to increase the readability and understandability of the specification.  
General ideas on the \focust modelling of components controlling behaviour of safety-critical systems in
their physical environment were introduced in \cite{spichkova2014modeling}. 
In this report we go further and provide a systematic review of the core features and operators. To discuss the general idea of these operators and features, we use a Steam Boiler System example to illustrate how the specification framework can be applied.

Alur and Dill~\cite{Alur94atheory,DecisionProblemsTA} introduced timed automata  
that are nowadays One of the most well-established models for the specification and
verification of real-time system design. 
Timed automata have many advantages and many application areas, but they assume perfect continuity of clocks which may not suit
to specification of 
embedded system with instantaneous reaction times. 
Timed automata  also do not prevent Zeno runs~\cite{ZenoRuns}:   an infinite number of transitions in a finite period of time cannot be excluded. This problem was solved in an extended version of timed automata presented \cite{Puri,Bouyer:2011}. %
\focust provides a completely different solution to this problem: the Zeno runs are excluded on the syntactical level. 
Here fe follow approach of Henzinger et al.~\cite{DigitalClocks} that   
any timed transition system can be discretised without loss of generality.
As \focust is based on \emph{Focus on Isabelle}, we can even switch from one time granularity to another using predefined operators.
 
 \textbf{Outline:} 
 The rest of the report is organised as follows.
 Section \ref{sec:types} presents the core data datypes and the notion of the timed \focust streams. 
 In Section \ref{sec:operators} we introduce the basic \focust operators as well as spation-temporal aspects of the language. 
Section \ref{sec:specification} presents the core structure of the \focust specifications.
In Section \ref{sec:features} we discuss a number of special features of the language that were introduced to increase its readability and understandability. 
Section \ref{sec:boiler} illustrates the presented ideas using an example specification of a  steam boiler system. 
Section \ref{sec:conclusions} summarises and concludes the report.
 
% 
%========================
\section{Data Types and Streams}
\label{sec:types}
 
 In both \Focus and \focust, the systems are built out of components, where the component  specifications are based on the notion of \emph{streams} that represent 
communication histories of \emph{directed channels}.. 
However, they have different syntax and semantics for streams:
\begin{itemize}
\item{\Focus:} 
Input and output streams of a component can be timed (taking into account timing aspects) or untimed (abstracting from all timing aspects). 
$M^{\underline{\omega}}$ denotes in \Focus the set of all 
timed streams, $M^{\underline{\infty}}$ and $M^{\underline{*}}$ denote the sets of 
all infinite and all finite timed streams over the set $M$ respectively.
Timed streams are mappings 
 of natural numbers $\Nat$ to the single messages, where
 a message can be either a data message of some type o \emph{time tick}
(represented by $\surd$):
\[ 
\begin{array}{l}
M^{\underline{\omega}}$ =  $M^{\underline{*}} \cup M^{\underline{\infty}}\\
%&&\\
M^{\underline{*}} \stackrel{\mathrm{def}}{=} 
\bigcup_{n \in \mathbb{N}}([1..n]\to M \cup \{ \surd \})\\
%&&\\
M^{\underline{\infty}} \stackrel{\mathrm{def}}{=} 
\mathbb{N}_+ \to M \cup \{ \surd \}
\end{array}
\]
\item{\focust:} 
Input and output streams of a component are always timed, as spatio-temporal aspects are the core of the framework. 
The (timed) streams are mappings from $\Nat$ to lists of messages  within the corresponding time intervals. 
Thus, these streams are infinite per default, but they could be empty completely or from a certain point which is represented by empty time intervals $\nempty$.  \\
More precisely, \focust has streams of two kinds:
	\begin{itemize}
	\item
	\emph{Infinite timed streams} are used to represent the input and the output streams;
	\item
	\emph{finite timed streams} are used  
	to argue about a timed stream that was truncated 
	at some point of time.
	\end{itemize} 
%\item
%
\end{itemize}

The base data types we use in \focust are \Bool, the type of truth values, 
\Nat, the type of natural numbers, and \nbit, the type of bit values 0 and 1.

Definitions of an enumeration and list types are inherited from the \Focus specification language, and can be represented
in two ways that have the same semantics:
\[
	\begin{array}{ll}
		\ntype~T = e_1 \mid \dots \mid e_n & \mbox{enumeration type} \\
		\ntype~T = \{e_1, \dots,  e_n\} & \mbox{enumeration type}  \\
		\ntype~L = \nfst{\Nat} & \mbox{list type over \Nat } 
	\end{array}
\]

The \focust \emph{records type} $RV$ is defined also using the \Focus rules, where $con_1, \dots, con_n$ and 
$sel_1^1, \dots, sel^n_{k_n}$ are constructors and selectors respecivelly:
\[
\begin{array}{rl}
\ntype~RV = & con_1(sel_1^1 \in T_1^1, \dots, sel^1_{k_1} \in T_{k_1}^1)\\
 & \dots \\
\mid       & con_n(sel_1^n \in T_1^n, \dots, sel^n_{k_n} \in T_{k_n}^n)
\end{array}
\]
Infinite timed streams of type $T$ are defined  by a functional type 
\[
\Nat\ \dto \nfst{T}
\] 
Finite timed streams of type  $T$  are defined by list of lists overt this type, i.e., 
\[
\nfst{(\nfst{T})}
\]
where $\nfst{T}$ denotes a list of elements of type $T$.

%========================
\section{Basic Operators}
\label{sec:operators}

The operator $\ti{s}{t}$ represents $t$th time interval of the stream $s$.

$\nempty$ denotes an empty list (an empty time interval).

$\angles{a_1, \dots, a_m}$ denotes a list of $m$ elements $a_1, \dots, a_m$.

The predicate $\msg{s}{k}$ holds if each time interval of the stream $s$ has at most $k$ elements. 
Thus, $\msg{s}{1}$ would mean that each time interval of $s$ either has a single element or is empty.

The predicate \nts{s} ensures that each time interval of the stream $s$ contains exactly one message. 
 Thus, we can say that \nts{s} implies $\msg{s}{1}$ but not vice versa.
 
We also allow to use standard logical quantifiers $\forall$ and $\exists$ as well as the following operators:
$\wedge$ denotes AND, $\vee$ denotes OR, $=$ denotes equality, $\to$ denotes implication.

To refine the time granularity, we use the operator \ntsplit{s}{n}. 
This operator splits every time interval of the stream $s$ into $n$ time intervals. 
We have defined three versions of this operator:
\begin{itemize}
	\item 
	to locate all messages from the original  time interval  to 
	the \emph{first} of the $n$ corresponding intervals,
		\item 
	to locate all messages from the original  time interval  to 
	the \emph{last} of the $n$ corresponding intervals,
  \item 
  to distribute 
  the messages from the time interval of the original stream over
 the $n$ corresponding intervals.
\end{itemize}

To make the time granularity more coarse, we use the operator \ntjoin{s}{n}. 
It joins $n$ time intervals of the stream $s$ into a single time interval.

The time stamp operator \utm\ returns for a timed stream $s$ and a natural number $k$ an 
index of time interval in which the $k$th message in the stream $s$ is transmitted. 

The filtering operator $\nfilter{M}{s}$ filters away messages from each time interval of the timed stream $s$ if these messages do not belong to the filtering set $M$.

To represent real objects that can physically change their location in space, we define so-called
\emph{sp-objects}.
An sp-object is defined not only by its behavioural specification but also by a tuple
\[
< location, speed, direction, radius, occupied space > 
\]
In \focust this tuple is specified using 
\begin{itemize}
\item
a special global (in the scope of the system specification) constant $rad$ associated with an elementary so-object to represent the radius of the maximal space
the sp-object can ``cover'' in the worst case; 
In the case an sp-object $S$ is a composition (system) of a number of other sp-objects, we calculate its
 $rad$  by analysing which space its subcomponents can occupy in the worst case:
\[
S.rad = max(WCX, WCY) / 2 
\]
 $WCX$ and $WCY$ being the maximum extensions of all of the subcomponents of $S$ in direction $x$ respective $y$;\\

\item
four
special global (in the scope of the system specification) variables to store for each sp-object its 
	\begin{itemize}
	\item
	current $location \in Space$ (i.e., central point of the sp-object),
	\item
	current $speed \in \Nat$, 
	\item
	current $direction \in Directions$, and
	\item
	current $rzone \in Zone$.
	\end{itemize}
\end{itemize}
The  type $Space$ is a tuple of two Cartesian coordinates $xx$ and $yy$ defined over \Nat:
\[ 
Space \ndef \Nat \times \Nat
\]   
The type $Directions$  represents an angle
in the Cartesian coordinate system:
\[
Directions \ndef \{ 0, \dots, 359 \}
\] 
The type $Zone$ is a tuple of Cartesian coordinates of two spatial points $X$ and $y$ $(minX, minY, maxX, maxY)$ defined over \Nat, where $X$ correspond to the upper left corner and $Y$ corresponds to the upper right corner of the corresponding zone.
\[ 
Space \ndef \Nat \times \Nat \times \Nat \times \Nat
\]  
The behavioural specification of the corresponding component can contain constrains on the speed, direction, and location of the so-object as well as on spatio-temporal dependencies among the so-objects in the system. 
While verifying the corresponding properties we can ensure, for example, that the object does not  exceed its speed limit, does enter specific areas or does not collide with another so-object.  

For composite so-objects we also have additional constraints:
{\footnotesize{$$
\begin{array}{r@{\;}l}
\forall S, & C: C \in subcomp(S) \to\\
& (S.rzone.minX \le S.C.rzone.minX \wedge S.rzone.minY \le S.C.rzone.minY) ~ \wedge\\
&  (S.rzone.maxX \ge S.C.rzone.maxX \wedge S.rzone.maxY \ge S.C.rzone.maxY) \\
&  ~\\
\forall k, & S, C: C \in subcomp(S) \to\\
& (k \le S.rzone.minX  \to (k + S.C.rad) \le S.C.location.xx)
  \end{array}
$$}}

%========================
\section{Core Structure of  \focust Specifications}
\label{sec:specification}

A template for a general \focust specification is presented below: 
 
 \begin{spec}{ComponentName}{timed}
\InOut{x_1 : InType_1, \dots, x_m : InType_m}{y_1 : OutType_1, \dots, y_n : OutType_n}
\tab{local} v_1 \in VarType_1, \dots, v_k  \in VarType_k\\
\tab{\uinit} v_1  = varValue_1, \dots, v_k = varValue_k
\zeddashline
\tab{\uasm}\\
\cbox{A1} \t1  FirstAssumptionFormula\\
\cbox{A2} \t1  SecondAssumptionFormula\\
\dots\\
\zeddashline
\tab{\ugar} \\
\cbox{I1} \t1 \ti{y_1}{0} = yValue_1\\
\cbox{I2} \t1 \ti{y_2}{0} = yValue_2\\
\dots\\
%
%\ \\
\forall t \in \Nat:\\
%1
\cbox{B1} \t1 FirstProperty 
\\
~\\
%2
\cbox{B2} \t1 SecondProperty\\
\\
~\\
%3
\cbox{B3} \t1 ThirdProperty 
\\ 
\dots
\\
\end{spec}

The \tbf{in} and \tbf{out} sections are used to specify input and output streams of the corresponding types:
\begin{itemize}
\item
$x_1,  \dots, x_m$ are input streams of the types $InType_1, \dots, InType_m$, respectively;
\item
$y_1, \dots, y_n$ are input streams of the types $OutType_1, \dots, OutType_n$, respectively.
\end{itemize}

The local variables and their initial values are specified in the sections \tbf{local} and \tbf{init}:
$v_1, \dots, v_k$   are local variables of the types $VarType_1, \dots,  VarType_k$ and with the 
initial values  $varValue_1, \dots, varValue_k$,
respectively.

We specify every component using assumption-guarantee-struc\-tu\-red templates.
This helps avoiding the omission of unnecessary assumptions about the system's environment since a specified component is required to fulfil the guarantee only if its environment behaves in accordance with the assumption.

The keyword \tbf{asm} lists the assumption that the specified component demands from its environment, for example  
that all the input streams should contain exactly one message per time interval (i.e., to be time-synchronous).

The component behaviour that should be guaranteed in the case all assumptions are fulfilled, is then described in the specification section \tbf{gar}. 
Each formula in the assumption and guarantee-section is numbered. 

For easier referencing, we propose to number assumptions by $A1, A2, A3, \dots$, 
initial guarantees by $I1, I2, I3, \dots$, 
and the core guaranteed behavioural properties by $B1, B2, B3, \dots$. 
The behavioural properties are usually either defined over all time intervals $t \in \Nat$ 
or are presented by the corresponding predicates, e.g., \uts.

Under the \emph{initial guarantees} we understand the initial values on the output streams:
in the case of strongly-causal components we might need to specify output values explicitly.

The guaranteed behaviour is specified as a special form of timed automata that we name \emph{Timed State Transition Diagrams} (TSTDs).
A TSTD can be described in as a diagram , a textual form, or 
a special kind of tables including a number of new operators that work on time intervals.

For a real-time system $S$ with a syntactic interface $\nint{I_S}{O_S}$, where
$I_S$ and $O_S$ are sets of timed input and output streams respectively,
a 
TSTD corresponds to a tuple
$%\[
(State, state_0, I_S,  O_S, \to)
$, %\]
in which
$State$ is a set of states, $state_0 \in State$ is the initial state, and
$\to \;\subseteq (State \times I_S \times State \times  O_S)$ represents the transition function of the TSTD.

%========================
\section{Readability and Usability}
\label{sec:features}

\focust\ allows to use so-called \emph{implicit else-case} constructs.
That means, if a variable is not listed in the guarantee part of a transition, it implicitly keeps its current value.
An output stream not mentioned in a transition will be empty.

In a component model, one often has transitions with local variables that are not changed.
Also, frequently outputs are not produced,
e.g., in the case when a component gets no input or
some preconditions necessary to produce a nonempty output are violated.
In many formal languages this kind of invariability has to be defined explicitly in order to avoid
underspecified component specifications.
To make our formal language better understandable for programmers, we use in
\focust\ so-called \emph{implicit else-case} constructs.
That means, if a variable is not listed in the guarantee part of a transition, it implicitly keeps its current value.
An output stream not mentioned in a transition will be empty.
 
We also do not require to introduce auxiliary variables explicitly:
The data type of a not introduced variable is universally quantified in the specification such that it can be used with any data value.

To increase readability of the model, we use the following colour notation:
\newpage
\begin{itemize}
\item
\tbf{Components:} strongly-causal elementary components are presented by blue blocks, weakly-causal elementary components are presented by green blocks, where the white blocks denote composite components;
\item 
\tbf{Streams:} the streams fulfilling the $\uts$ property are marked red, the streams fulfilling the $\msg{}{1}$ property are marked blue, all other streams are marked black.
\end{itemize}

%========================
\section{Example: Steam Boiler}
\label{sec:boiler}

The main idea of the steam boiler specification was taken from \cite{focus}. 
The steam boiler has a water tank, which contains a number of gallons of water, and
a pump, which adds $10$ gallons of water per time unit to its water tank, 
if the pump is on. At most $10$ gallons of water are consumed per time unit by
the steam production, if the pump is off.
The steam boiler has a sensor that measures the water level. 
Initially,  the water level is $500$ gallons, and the pump is off.

The specification group \emph{SteamBoiler} consists of the following components: 
\emph{SystemReq} (general requirements specification), 
\emph{ControlSystemArch} (system architecture),  
\emph{SteamBoiler}, \emph{Converter}, and \emph{Controller}. 
\\
We define the data type \emph{WaterPumpState} to denote the state of the steam boiler pump:
\[
\ntype~ WaterPumpState = PumpOn \mid PumpOff \\
\]

The specification \emph{SystemReq} describes  the requirements for the steam boiler system: 
(1) in each time interval the system outputs it current water level in gallons
 and this level should always be between $200$ and $800$ gallons; 
 (2) the system outputs the information on the water level each time interval. 

The specification \emph{Controller} describes the controller component of the system. 
The controller is responsible for switching the steam boiler pump on and off, and it remembers the current state of the pump as its local state. 
The behaviour of this component is asynchronous  to keep
the number of control signals (to switch the pump on and off) as small as possible.
It is weakly-causal, having no delays in his output.

{\footnotesize
\begin{spec}{\spc{SystemReq}}{td}
\Out{currentWaterLevel : \nat}
\tab{\uasm}\\
\cbox{A1} \t1  \ntrue\\
\zeddashline
\tab{\ugar} \\
\cbox{B1} \t1  \nts{currentWaterLevel}\\
\cbox{B2} \t1  \forall t \in \nat: 200 \le \nft{\ti{currentWaterLevel}{j}} \le 800
\end{spec} 
}

{\footnotesize
\begin{spec}{\spc{ControlSystemArch}}{td}
~\\
\centering
\includegraphics[width=9cm]{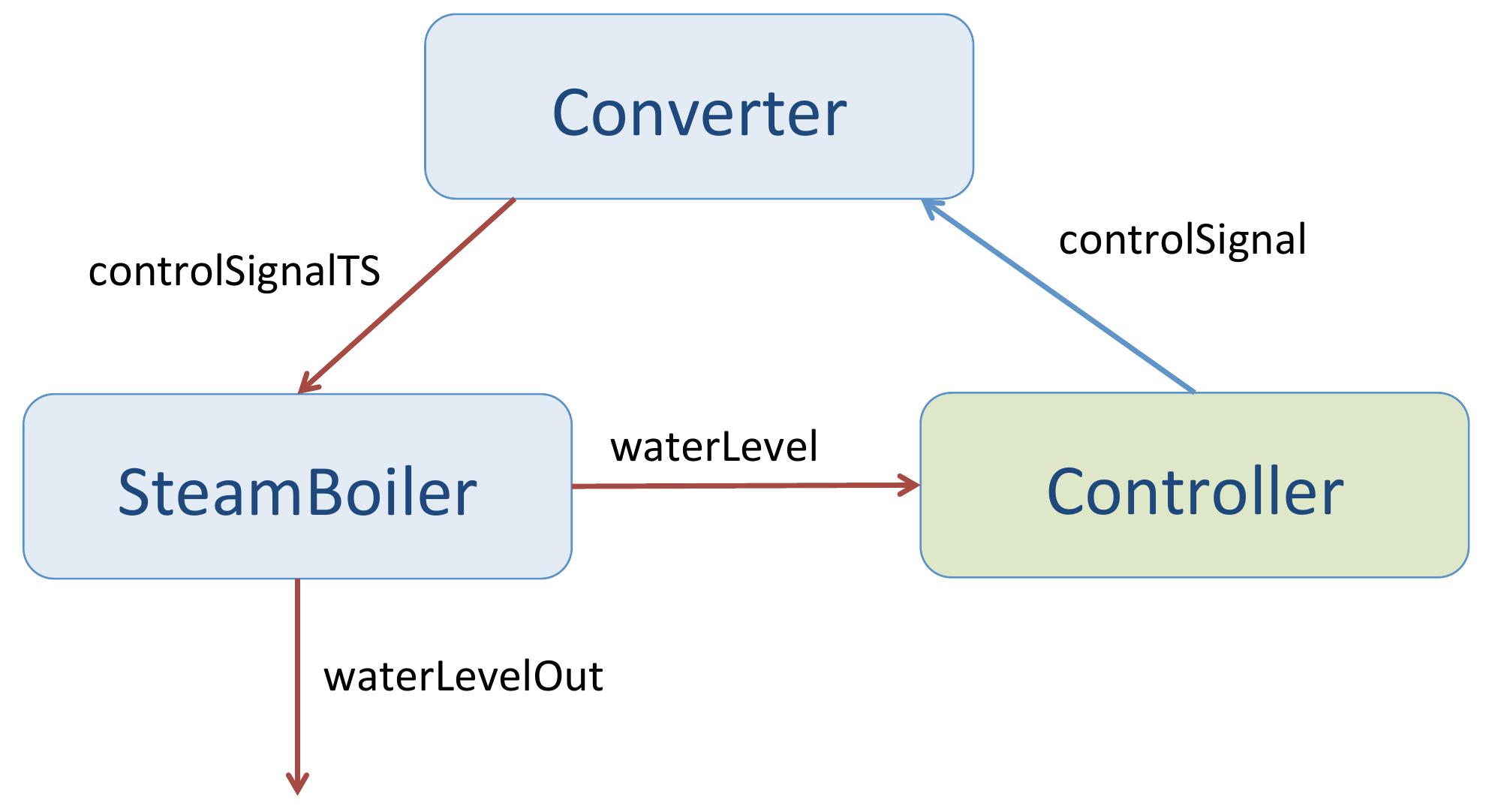}
\end{spec}
}

\begin{spec}{Controller}{timed}
\InOut{waterLevel : \Nat}{controlSignal: \Bit}
\tab{local} pump \in WaterPumpState\\
\tab{\uinit} pump = PumpOff
\zeddashline
\tab{\uasm}\\
\cbox{A1} \t1  \nts{waterLevel}\\
\zeddashline
\tab{\ugar} \\
% 
%
%\ \\
\forall t \in \Nat:\\
%1
\cbox{B1} \t1 pump = PumpOff ~\wedge~ \nft{\ti{waterLevel}{t}} > 300 \to \\
\t3 pump' = PumpOff  ~\wedge~ \ti{controlSignal}{t} = \nempty 
\\
~\\
%2
\cbox{B2} \t1 pump = PumpOff ~\wedge~ \nft{\ti{waterLevel}{t}} \le 300 \to \\
\t3 pump' = PumpOn  ~\wedge~ \ti{controlSignal}{t} = \angles{1}\\
\\
~\\
%3
\cbox{B3} \t1 pump = PumpOn~\wedge~ \nft{\ti{waterLevel}{t}} < 700 \to \\
\t3 pump' = PumpOn ~\wedge~ \ti{controlSignal}{t} = \nempty 
\\
~\\
%4
\cbox{B4} \t1 pump = PumpOn~\wedge~ \nft{\ti{waterLevel}{t}} \ge 700 \to \\
\t3 pump' = PumpOff ~\wedge~ \ti{controlSignal}{t} = \angles{0}
\\
\end{spec}

 The original \Focus specification of the $Controller$ component, cf. \cite{focus}, 
uses mutually recursive functions \emph{on} and \emph{off} 
to specify the local state of the component. To increase the understandability of the specification, 
we simply use the component state instead of mutually recursive functions.

In this particular case, 
we can also provide a semantically equivalent specification of controller that consist of  a single formula 
combined out of if-then-else-fi constructs:
 
 \begin{spec}{Controller}{timed}
\InOut{waterLevel : \Nat}{controlSignal: \Bit}
\tab{local} pump \in WaterPumpState\\
\tab{\uinit} pump = PumpOff
\zeddashline
\tab{\uasm}\\
\cbox{A1} \t1  \nts{waterLevel}\\
\zeddashline
\tab{\ugar} \\
% 
%
%\ \\
\forall t \in \Nat:\\
%1
\cbox{B1} \t1 \nif pump = PumpOff \\
\t2 \nthen  \nif \nft{\ti{waterLevel}{t}} > 300 \\
	\t4 \nthen  pump' = PumpOff  ~\wedge~ \ti{controlSignal}{t} = \nempty  \\
	\t4 \nelse pump' = PumpOn  ~\wedge~ \ti{controlSignal}{t} = \angles{1} \\
	\t4 \nfi \\
\t2 \nelse \nif  \nft{\ti{waterLevel}{t}} < 700 \\
	\t4 \nthen pump' = PumpOn ~\wedge~ \ti{controlSignal}{t} = \nempty \\
	\t4 \nelse pump' = PumpOff ~\wedge~ \ti{controlSignal}{t} = \angles{0}\\
	\t4  \nfi \\
\t2 \nfi
\\ 
\end{spec}

 The specification \emph{SteamBoiler} describes steam boiler component, which has to control  
the current water level  every time interval.
The initial water level is specified to be $500$ gallons.
For every point of time the following must be true: 
\begin{itemize}
\item
if the pump is off, 
the boiler consumes at most $10$ gallons of water (i.e., any number of gallons between 0 and 10), 
\item
if the pump is on, at most $10$ gallons of water (i.e., any number of gallons between 0 and 10) will be added to its water tank.
\end{itemize}

The \emph{Converter} component simply converts the asynchronous output produced by the controller
to time-synchronous input for the steam boiler. 

 \begin{spec}{SteamBoiler}{timed}
\InOut{controlSignalTS : \Bit}{waterLevel, waterLevelOut: \Nat}
\tab{\uasm}\\
\cbox{A1} \t1  \nts{controlSignalTS}\\
\zeddashline
\tab{\ugar} \\
\cbox{I1} \t1 \ti{waterLevel}{0} = \angles{500}\\
~\\ 
%1
\cbox{B1} \t1 waterLevel = waterLevelOut
\\
~\\
\forall t \in \Nat:\\
%2
\cbox{B2} \t1 \exists   r \in \Nat: 0 < r \le 10 ~\wedge~\\
          \t6  \nif \ti{controlSignalTS}{t} = \angles{0} \\
          \t6 \nthen \ti{waterLevel}{t+1} = \angles{\nft{\ti{waterLevel}{t}} - r}\\
          \t6 \nelse \ti{waterLevel}{t+1} = \angles{\nft{\ti{waterLevel}{t}} + r}\\
          \t6 \nfi 
\\ 
\end{spec}

\begin{spec}{Converter}{timed}
\InOut{controlSignal : \Bit}{controlSignalTS: \Bit}
\tab{local} currentControlSignal \in \Bit\\
\tab{\uinit} currentControlSignal = 0
\zeddashline
\tab{\uasm}\\
\cbox{A1} \t1  \msg{1}{controlSignal}\\
\zeddashline
\tab{\ugar} \\
\cbox{B1} \t1 \nts{controlSignalTS}\\
~ \\
\forall t \in \Nat:\\
%1
\cbox{B2} \t1 \nif \ti{controlSignal}{t} \neq \nempty \\
		\t5 \nthen currentControlSignal' = \nft{\ti{controlSignal}{t}}\\
		\t5 \nelse currentControlSignal' = currentControlSignal\\
		\t5 \nfi \\
~\\
\cbox{B3} \t1 \ti{controlSignalTS}{t} = \angles{currentControlSignal}
\end{spec}

%==================
\section{Conclusions}
\label{sec:conclusions}

The  understandability, comprehensibility and scalability of the formal specifications have been hypothesized as hindering factors for their adoption in industry.  
Jones et~al.  \cite{jones1996formal}, Jackson \cite{jackson2001lightweight}, Atzeni et~al. \cite{atzeni2014lightweight} as well as 
Bennion and Habli \cite{bennion2014candid} presented promising approaches  in the context of industrial projects: lightweight formal methods, where the lightweight verification does not require special expert and can be assigned to the testing group. To make a formal method really adopted, it should be not only sound, but also comprehensive and easy-to-understand. 

In this report we presented a summary of the  core operators and features of the \focust specification framework. 
The goal of this summary is to provide a basis for further spatio-temporal analysis of the system properties as well as an interface for the further connection of the \focust\ to other frameworks and techniques.  
The framework was developed with a special attention to 
 the human factor analysis within formal methods  to increase the readability and understandability of the formal specifications.
  To discuss the general idea of these operators and features and  to illustrate how the specification framework can be applied, we use a Steam Boiler System example, which is one of the common examples to present interactive systems.
   
Our future research direction comprises work on the  modelling levels for spatio-temporal systems, that reflect the idea of
remote integration/interoperability testing in a virtual environment \cite{enase2014,enase2014book,issec2013spichkova}, 
as well as on the optimisation of the verification methodology focusing on the spatial aspects and the corresponding case studies.

%\newpage
%\bibliographystyle{abbrv}
%\bibliography{biblio}

\end{document}